\documentstyle[11pt,epsfig,astron]{article}
\begin{document}
\begin{titlepage}
\title{The inverse problem in microlensing : from the optical depth to the galaxy models parameters}

 \author{{V. F. Cardone$^1$, R. de Ritis$^{1,2}$, A. A. Marino$^3$} \\ 
{\em \small $^1$Dipartimento di Scienze Fisiche, Universit\`{a} di 
Napoli,} \\
{\em \small Mostra d'Oltremare pad. 20-80125 Napoli, Italy;} \\
{\em \small $^2$Istituto Nazionale di Fisica Nucleare, Sezione di Napoli,} \\
{\em \small Complesso Universitario di Monte S. Angelo, Via Cinzia, Edificio G 80126 Napoli, Italy;} \\
{\em \small $^3$Osservatorio Astronomico di Capodimonte,} \\
{\em \small Via Moiariello, 16-80131 Napoli, Italy.}}
	      \date{}
	      \maketitle
	      \begin{abstract}

We present in this paper a simple method to obtain informations on galaxy models parameters using the measured value of the microlensing optical depth. Assuming a 100\% MACHO's dark halo, we ask the predicted optical depth for a given model to be the same as the observed one, in a given direction. Writing the optical depth in terms of the given halo model parameters and inverting this relation with respect to one of them, it is possible to get information on it, fixing under reasonable hypothesis the other parameters. This is what we call the {\it inverse problem in microlensing}. We apply this technique to the class of power\,-\,law models with flat rotation curves, determining the range for the core radius $R_c$ compatible with the constraints on the halo flattening $q$ and the measures of $\tau$ towards LMC. Next, we apply the same method to a simple triaxial model, evaluating the axial ratios.

	      \end{abstract}

\vspace{20. mm}

e-mail addresses: \\
winny@na.infn.it \\
deritis@na.infn.it \\
marino@na.astro.it \\
	      \vfill
	      \end{titlepage}

\section{Introduction}

Gravitational microlensing is a powerful tool to investigate the dark halo, whose presence around galaxies is suggested by the flatness of rotation curves. The discovery of microlensing events towards LMC and SMC has witnessed the presence of MACHO's as baryonic costituents of the dark halo, but has also opened many new problems. What is the fraction $f$ of dark matter composed by MACHO's? How are they distribuited? And (last but not least) what is the shape of the dark halo?
 
In these years several authors have tried to answer these questions using the data coming from microlensing observations towards Magellanic Clouds to learn something about the properties of the dark halo. To start the investigations, Paczynski (1986) has adopted the most simple model: the singular isothermal sphere. Few years later, Griest (1991) suggested to add a core radius to eliminate the central singularity of the mass density of Paczynski' s model, obtaining what is now called {\it the standard model}. Unfortunately, optical depth and event rate so predicted are higher than observed, therefore many authors have concluded that MACHO's are not the only components of the dark halo. A simple way to estimate $f$ is to do the ratio between the observed optical depth and the predicted value for a model totally made of MACHO's. However, this method makes $f$ model dependent, i.e. $f$ depends on the model adopted to describe the shape of the halo and the distribution of MACHO's inside it. In principle, there is no reason a priori to say that MACHO's are not the only components of the dark halo. In this paper we consider $f = 1$ and show how it's possible to use the measured value of the optical depth to extract informations on galaxy parameters. Instead of going from galactic dynamics to microlensing, we try the other way around, that is going from microlensing to galactic dynamics. We will express the optical depth $\tau$ as function of some galaxy model parameters $(x, y, ...)$ and then invert the relation so found. If $f= 1$, then $\tau_{theor}$, i.e. the value predicted for a 100\% MACHO's halo, has to be equal to the measured value $\tau_{obs}$; so, we may use the function $\tau = \tau(x, y, ...)$ to extract informations on $(x, y, ...)$. This is what we call the {\it inverse problem in microlensing}
\footnote{A similar technique has been yet used in other works ( see e.g. De Paolis, Ingrosso \& Jetzer, 1996), but it has never been applied to the models we consider here.}.

In this paper, we apply this method of investigation to two classes of galaxy models and try to learn something about their characteristics. We do not consider spherical models since there are few evidences which lead to reject this extremely simple shape for the dark halo. In fact, it is possible to estimate the halo flattening through dynamical modelling of collissionless kinematic tracers, such as extreme Poulation II stars. These probe the potential at large distances and high above the disk and permit to estimate the ratio $c/a$ between the principal semiaxes, which turn out to be $(c/a)_{trc} = 0.6 - 0.85$ \cite{SZ90}. Using the Jeans equations, van der Marel (1991) has shown that halo flattenings of $0.4 \le (c/a)_{halo} \le 1$ were consistent with the data depending on the tilt of the velocity ellipsoid away from the disk plane. These evidences suggest to give off the spherical simmetry. A step in this direction is the adoption of axisimmetric model for the dark halo, this is the reason why many authors have used the so called {\it power\,-\,law models} \cite{E93}, \cite{E94}
\footnote{However, flattened models, other than power\,-\,law, has been also considered \cite{Jetz91}.}. 
Optical depth and event rate towards LMC have been calculated \cite{EJ94}, \cite{A95} and they found a method to experimentally estimate the halo flattening from the ratio between event rate observed towards LMC and SMC. Unfortunately, predicted and observed optical depth are once again not consistent. In this paper, we apply the inverse problem method to the class of power\,-\,law models with flat rotation curves and try to learn something about their properties to see whether it is possible to explain the disagreement yet found between theory and observations of microlensing events towards LMC.

It is worthy to note, however, that there are no direct evidences of deviations of the Milky Way halo from axisymmetry in the disk plane. This may be the consequence of a partial compensation between disk and halo potentials, in case they are of comparable dynamical importance. It happens infact that, even for significantly triaxial halos, the combined disk\,-\,halo potential is much more symmetric than the single components potentials \cite{R95}.
Then, it is not clear whether one must give off the axisymmetric hypothesis. An help in this choice may come from the theoretical predictions for the halos arising in hierarchical structure formation scenarios. For non dissipational dark matter, this question has been addressed by cosmological N\,-\,body simulations \cite{DC91}, \cite{KG92}, \cite{NFW96}. Despite the ongoing discussion about numerical simulations and initial conditions, all dissipationless simulations agree on predicting a strongly triaxial shapes, closer to prolate than oblate. The resulting axis ratios vary slowly with radius and the distribution of halo flattenings ($c/a$) peaks near 0.7. However, dissipational dark matter (such as MACHO's and cold molecolar clouds) will affect the halo shape in different ways, compressing the halo in the disk plane direction. Simulations including gas dynamics \cite{KG92} or a slow growth of a disk component \cite{D94} have found that halos become slightly flatter and much closer to oblate and that these changes extend to large radii. Our position in the Milky Way complicates measurements of the halo shape so that many models are consistent with the data coming from different kind of measurements (for a brief review see \cite{R95}). Better results have been obtained for external disk galaxies. In particular, flattening estimates are possible for several polar ring galaxies ( see e.g. \cite{SS90}), whose halos are definetely flattened perpendicular to the stellar disk according to numerical simulations. If our dark halo is typical, then cosmological simulations and observations around other spiral galaxies suggest a triaxial halo, flattened perpendicular to the disk plane. A result in this sense has been obtained by Binney, May \& Ostriker (1987). They constructed approximately self\,-\,consistent disk halo models with Staeckel potentials and find that $(c/a)_{halo} \sim 0.5$ fits the observations best.

Following the suggestions coming from numerical simulations and the cited result of Binney, May \& Ostriker, in this paper we also adopt a particular separable triaxial model to describe the dark halo and evaluate the predicted optical depth towards LMC. Doing this requires the knowledge of the axial ratios and the extension $D = a_1$ of the model, which are unknown. However, we will show how our method may estimate the axial ratios under the hypotheses $f = 1$ and $D = 100 \ {\rm kpc}$. 

In Section 2, we briefly review some basics notions of microlensing, introducing the optical depth $\tau$ and the event rate $\Gamma$. Section 3 is devoted to the description of power\,-\,law models, while in Section 4 we calculate the optical depth towards LMC and apply the method of the inverse problem. Section 5 is dedicated to the discussion of the results obtained for the power\,-\,law models with flat rotation curves. The perfect ellipsoid, a particularly simple separable triaxial model, is described in Section 6, while in the following Section 7 we do the same analysis as in Section 4 for this model. The relative results are summarised in Section 8, while the next section is devoted to conclusions.

\section{Basics of microlensing}

When a lens moves across the line of sight, the light coming from the source is amplified by a factor \cite{MR97}, \cite{J98}:
\begin{equation}
A(u) = \frac{u^2 + 2}{u \sqrt{u^2 + 4}}
\label{eq: uno}
\end{equation}
where $u = b/R_e$ is the impact parameter in units of Einstein radius :
\begin{equation}
R_e = \sqrt{\frac{4GM D_d ( D_s - D_d )}{c^2 D_s}} .
\label{eq: due}
\end{equation}
Here, $D_s$ is the distance of the source, $D_d$ is the distance along the line of sight and $M$ is the mass of the lens. If the MACHO is moving with a transverse velocity $v_{\perp}$, the impact parameter is time dependent and is given by :
\begin{equation}
u(t) = \sqrt{u^2_{min} + \omega^2 ( t - t_0 )^2}
\label{eq: tre}
\end{equation}
being $u_{min} = u( A_{max} )$ and $\omega = v_{\perp}/R_e$. Experimentally microlensing events are characterized by the maximum amplification $A_{max}$ and the event duration $T = 1/\omega$.

Two are the most important parameters in microlensing analyses : the {\it optical depth} and the {\it event rate}. According to Vietri \& Ostriker (1983), the optical depth $\tau$ is the probability of observing a microlensing event. In the little number limit, this is just the number of MACHO's inside the microlensing tube, a cylinder whose axis is the line of sight with radius $u_{th}R_e$, being $u_{th} = u(A_{th})$ with $A_{th}$ the threshold amplification \cite{Pac86}. When $A_{th} = 1.34$, $u_{th} = 1$. The optical depth is then \cite{Pac86}, \cite{J98}:

\begin{equation}
\tau = \pi \int{\frac{1}{M}\rho R_e^2 ds}
\label{eq: quattro}
\end{equation}
where we have posed $s = D_d/D_s$.

The event rate $\Gamma$ is simply the number of events divided by the duration time of observation. Analitically, it is defined as \cite{DeR91} : 
\begin{equation}
\Gamma = \int{\varepsilon(T) \frac{d\Gamma}{dT} dT}
\label{eq: cinque}
\end{equation}
where $\varepsilon(T)$ is the efficiency function and $d\Gamma/dT$ is the differential rate; in the hypothesis of same mass, it is given by :
\begin{equation}
\frac{d\Gamma}{dT} = 2u_{th} \int_{0}^{2\pi}{d\gamma}
\int_0^{D_s}{dD_d \left ( \frac{R_e}{T} \right )^4}
\int{dv_{\parallel} f({\bf x},{\bf v})}
\label{eq: sei}
\end{equation}
being {\it f({\bf x},{\bf v})} the distribution function\footnote{Remember that the distribution function (hereafter DF) of a collisionless system is the numerical density on the configuration space, i.e. the number of particles with position in the interval $({\bf x}, {\bf x} + d^3x)$ and velocity in $({\bf v}, {\bf v} + d^3v)$ \cite{BT87}.} 
 of the model adopted for describing the distribution of MACHO's and $v_{\parallel}$ is the MACHO velocity parallel to the line of sight.

Optical depth, event duration and event rate are related by the useful relation \cite{DeR91}, \cite{J98} :
\begin{equation}
< T > = \frac{2 u_{th} \tau}{\pi \Gamma}
\label{eq: sette}
\end{equation}
where we have defined :
\begin{equation}
< T > = \frac{1}{\Gamma} \int{T \frac{d\Gamma}{dT} dT}.
\label{eq: settebis}
\end{equation}
Eq. (\ref{eq: sette}) is similar to the one used to estimate optical depth from the observed events. The latter is :
\begin{equation}
\tau_{obs} = \frac{\pi}{2N_{\star} t_{obs}} \Sigma_i
\frac{T_i}{\epsilon(T_i)}
\label{eq: tauobs}
\end{equation}
where the sum is on the events observed, being $N_{\star}$ the number of resolved monitored stars, $t_{obs}$ the duration of observations and $\epsilon(T_i)$ the efficiency for the i\,-\,th event.

\section{Power\,-\,law models}

The event rate is not easy to calculate since the DF $f({\bf x}, {\bf v})$ cannot be prescribed arbitrarly. Under reasonable hypotheses, the halo is a collisionless system and its DF must satisfy {\it collisionless Boltzmann equation} \cite{BT87}. By Jeans' theorems, the DF depends on configuration space coordinates only through the three (or just one or two of them) isolating integrals of motion in the gravitational potential generated by the galaxy itself. So the construction of self consistent galactic models is not so easy, especially for non spherical systems.  

We start to dedicate our attention to a class of axisymmetric galactic halo models\,: the {\it power\,-\,law models}, also known as Evans models \cite{E93} \cite{E94}. In this case the two integrals of motion are the energy {\it E} and the z\,-\,component $L_z$ of angular momentum. The DF is :
\begin{displaymath}
f( E, L_z ) = AL_z^2 E^{4/\beta - 3/2} + BE^{4/\beta - 1/2} + CE^{2/\beta - 1/2} \ \ \ \ ({\rm for} \ \beta > 0) \ ;  
\end{displaymath}  
\begin{displaymath}
f( E, L_z ) = AL_z^2 \exp{4E/v_0^2} + B \exp{4E/v_0^2} +C \exp{2E/v_0^2}
\ \ \ \ ({\rm for} \ \beta = 0) \ ;
\end{displaymath}
\begin{equation}
f( E, L_z ) = \frac{AL_z^2}{(-E)^{3/2 - 4/\beta}}
+ \frac{B}{(-E)^{1/2 - 4/\beta}} +
\frac{C}{(-E)^{1/2 - 2/\beta}} \ \ \ \ ({\rm for} \ \beta < 0) 
\label{eq: otto}
\end{equation}
where A, B, C are constants given in \cite{E94}.

Integrating the DF on the velocity space, one obtains the mass density of the models, which is all we need to evaluate the optical depth :
\begin{equation}
\rho( R, z ) = \frac{v_0^2 R_c^{\beta}}{4 \pi G q^2}
\frac{R_c^2 ( 1 + 2q^2 ) + R^2( 1 - \beta q^2 ) + z^2[ 2 - q^-2(1+\beta)]}
{(R_c^2 + R^2 + z^2q^{-2})^{\frac{\beta + 4}{2}}} 
\label{eq: nove}
\end{equation}
where $(R,z)$ are the usual cylindrical coordinates. The rotation curve is given by :
\begin{equation}
v_c^2(R) = \frac{v_0^2 R_c^{\beta} R^2}{(R_c^2 + R^2)^{\frac{\beta + 2}{2}}}
\label{eq: dieci}
\end{equation}

The parameters of the power\,-\,law models are the following :
\begin{enumerate}
\item{The core radius $R_c$, wich measures the scale at which the density begins to soften. Its value is not well known; Bachall, Schmidt \& Soneira (1983) give an estimate of $R_c$ as 2 kpc from star count data, while Caldwell \& Ostriker (1981) state $R_c = 10$ kpc. However, values as large as 20 kpc are also possible if we accept the hypothesis of maximal disk, i.e. rising the total mass of the disk and its contribution to the gravitational potential of the whole Galaxy.}

\item{The parameter $\beta$, which determines the asymptotic behaviour of the rotation curve; in fact, it is defined as :
\begin{equation}
\beta = - {\rm lim_{R \rightarrow \infty}} \frac{{\rm d} \log{v_c^2(R)}}{{\rm d}R}
\label{eq: undici}
\end{equation}
The value of $\beta$ is not known since it is not possible to measure the star rotation curves of the Galaxy beyond 20 kpc. However, the data seem to suggest that Milky Way is similar to other spiral galaxies, whose rotation curve is flat ($\beta = 0$) or only slowly rising($\beta < 0$). So, in the following we confine our attention to the case $\beta = 0$ and try to learn something on the other parameters from the analyses of the present measure of optical depth ( see later, Section 4).}

\item{$q$ describes the flattening of the dark halo : actually it is the axial ratio of the equipotential spheroids; $q = 1$ is the value for a spherical halo, while $q = 0.7$ is for an E6 halo \cite{E94}. The isophotal ellipticity is a function of $q$ and of the other parameters of the model. Obviously, no visible constraints are known for the dark halo; so we have to examine the whole physically possible range for {\it q}. This is fixed by the requirement that the mass density and the DF are always not negative. If we choose $\beta = 0$, these conditions define the range \cite{E93} :
\begin{displaymath}
0.71 \le q \le 1.08.
\end{displaymath}
}

\item{The distance $R_0$ of the Sun from the Galactic centre. Recently, this quantity has been reviewed by Reid (1989), who found that most recent determinations lie between 7 and 9 kpc; we will use the value 8 kpc, but we also consider the other values.}

\item{The normalization velocity $v_0$ which determines the typical velocity of MACHO's in the halo. This value is fixed by the local circular velocity, $v_c(R_0)$, through the formula (\ref{eq: dieci}). In our case ($\beta = 0$) we have :
\begin{equation}
v_0^2 = \frac{R_c^2 + R_0^2}{R_0^2} v_c^2(R_0)
\label{eq: dodici}
\end{equation}
We have to choose a value for $v_c(R_0)$; following Merrifield (1992), we adopt $v_c(R_0) = 200 \pm 10$ km/s, even if the IAU value is 220 km/s \cite{KL86}.}

\end{enumerate}

\section{The optical depth for power\,-\,law models with flat rotation curves}

To estimate the optical depth as in eq. (\ref{eq: quattro}), we need just the mass density of the models, which is given in (\ref{eq: nove}). Changing coordinates from cylindrical to heliocentric and executing the integral, one obtains \cite{EJ94}\,,\,\cite{A95}:
\begin{equation}
\tau = \frac{v_0^2 R_c^{\beta} u_{th}^2}{c^2 q^2 D_s^{\beta}}
\int_0^1{\frac{s ( 1 - s )(A_{\tau} s^2 + B_{\tau}s + C_{\tau})^2}
{(D_{\tau}s^2 + E_{\tau}s + F_{\tau})^{\frac{\beta +4}{2}}} ds}
\label{eq: tredici}
\end{equation}
where the constants $A_{\tau}, B_{\tau}, C_{\tau}, D_{\tau}, E_{\tau}, F_{\tau}$ are given in eq. A7 of \cite{A95}. For the class of power\,-\,law models with $\beta = 0$, they are :
\begin{displaymath}
A_{\tau} = \cos^2{b} + ( 2 - q^{-2} ) \sin^2{b} ;
\end{displaymath}

\begin{displaymath}
B_{\tau} = - 2 D_s^{-1} \cos{b}\cos{l} ;
\end{displaymath}

\begin{displaymath} 
C_{\tau} = [ R_0^2 + R_c^2 ( 1 + 2q^{-2} ) ] / D_s^2 ;
\end{displaymath} 

\begin{displaymath}
D_{\tau} = \cos^2{b} + q^{-2} \sin^2{b} ;
\end{displaymath}

\begin{displaymath}
E_{\tau} = - 2R_0 D_s^{-1} \cos{b} \cos{l} ;
\end{displaymath}

\begin{displaymath}
F_{\tau} = ( R_0^2 + R_c^2 )/ D_s^2 .
\end{displaymath}

The optical depth is then easy to evaluate in this case (in which $\beta = 0$ and $u_{th} = 1$) and the result is :
\begin{equation}
\tau = \frac{v_0^2 I}{c^2 q^2}
\label{eq: quattordici}
\end{equation}
being {\it I} the integral in eq. (\ref{eq: tredici}).

In fig. 1a we report the optical depth towards LMC\footnote{$D_s = 50$ kpc; $(b, l) = (-32^{\circ}.9, 280^{\circ}.5) $.}
in function of the core radius $R_c$ and the flattening parameter $q$, having fixed : $\beta = 0$, $R_0 = 8$ kpc, $v_c(R_0) = 200$ km/s.

\begin{figure}[ht]
\centering
\mbox{ \epsfxsize=13cm \epsfysize=10cm \epsffile{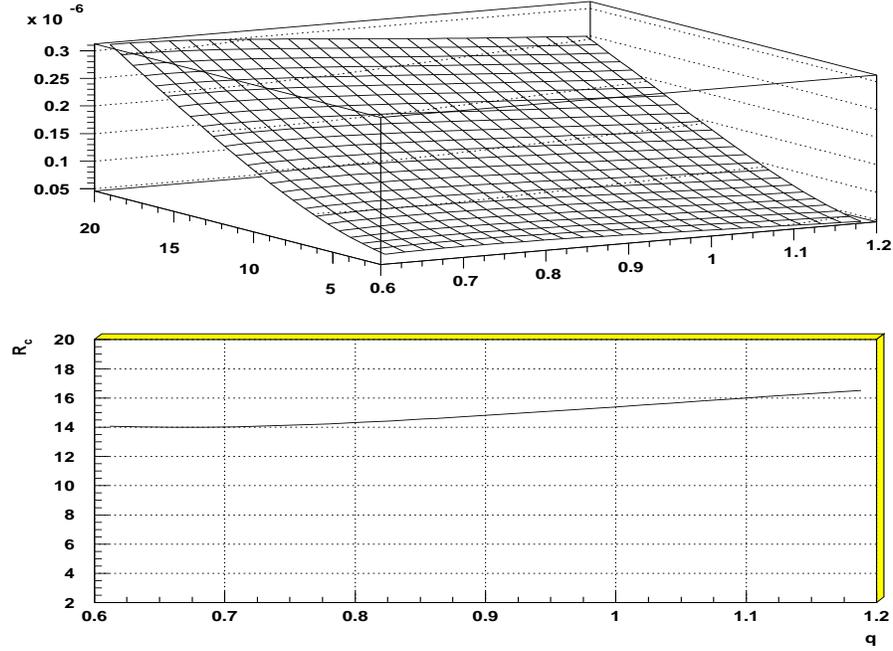}}
\caption{a.) Optical depth $\tau$ \ vs. \ $q$ \ and \ $R_c$ for power\,-\,law models with $\beta = 0$, $R_0 = 8$\,kpc, $v_c(R_0) = 200$\,km/s; b.) section of a) for $\tau = 2.1 \times 10^{-7}$.}
\end{figure}

Following the idea described in the introduction, we use this graph to extract information about the parameters of this class of power\,-\,law models, trying to learn something about the shape of the halo or the asymptotic behaviour of the rotation curve. The measured value of the optical depth towards LMC is : $\tau_{obs} = 2.1_{-0.8}^{+1.3} \times 10^{-7}$ \cite{B98}. We consider then a section of fig. 1a at the level $\tau = 2.1 \times 10^{-7}$ and obtain the graph in fig. 1b.

\begin{figure}[ht]
\centering
\mbox{ \epsfxsize=13cm \epsfysize=10cm \epsffile{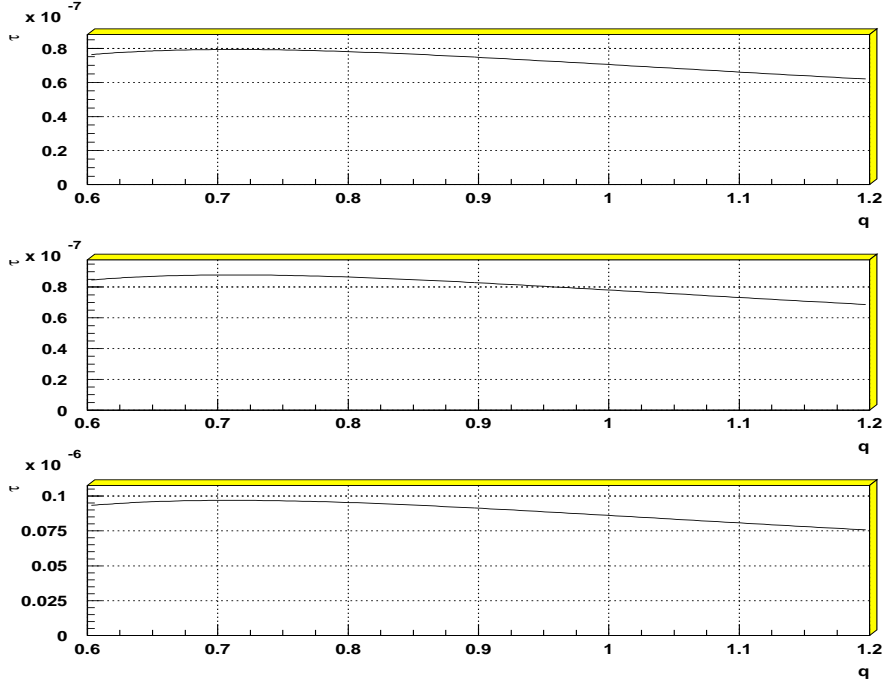}}
\caption{Optical depth $\tau$ (towards LMC) vs. halo flattening $q$, having fixed $\beta = 0$, $R_c$ = 5.6 kpc; $R_0 = 8$\,kpc, $v_c(R_0) = 190, 200, 210$\,km/s.}
\end{figure}

For every value of $q$ in the physically possible range, there is only one value of the core radius $R_c$ for which the model ( with the specified value of q and the other parameters fixed as before ) reproduces an optical depth equal to the one observed. In particular, the range for the core radius turns out to be :
\begin{equation}
14.03 \ {\rm kpc} \le R_c \le 15.91 \ {\rm kpc}
\label{eq: quindici}
\end{equation}

This may be compared to the range delimited by observations, which is 2 $\div$\,10\,kpc; excluding the hypothesis of maximal disk, the two ranges do not overlap.

However, one must also consider different choices for $v_c(R_0)$ and $R_0$ other than 200 km/s and 8.0 kpc. In fact, $\tau$ is strongly dependent on this two parameters. In fig. 2, we report the graph of $\tau$ vs. $q$ for $v_c(R_0) = 190, 200, 210$ km/s, having fixed $\beta = 0$, $R_c = 5.6$ kpc \cite{J98}, $R_0 = 8.0$ kpc.

The relative variation of $\tau$ (for $q$ = 0.8) is :
\begin{displaymath}
\Delta \tau_v( q = 0.8 ) = \frac{\tau( v_c(R_0) = 210 ) - \tau(v_c(R_0) = 190 )}
{\tau( v_c(R_0) = 210 )} \simeq 20 \% .
\end{displaymath}

Fig. 3 is, instead, a graph of $\tau$ vs. $q$ for $R_0 = 7, 8, 9$ kpc, being $v_c(R_0) = 200$ km/ s and $\beta$ and $R_c$ as before. The relative variation is :
\begin{displaymath}
\Delta \tau_r( q = 0.8 ) = \frac{\tau(R_0 = 7) - \tau(R_0 = 9)}
{\tau(R_0 = 7)} \simeq 29 \% .
\end{displaymath}

\begin{figure}[ht]
\centering
\mbox{ \epsfxsize=13cm \epsfysize=10cm \epsffile{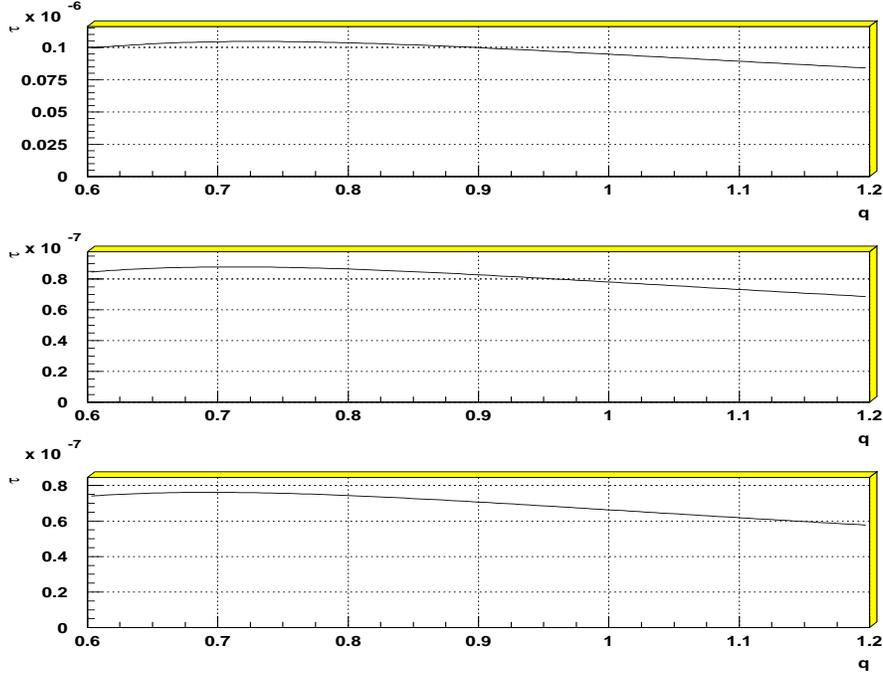}}
\caption{Optical depth $\tau$ (towards LMC) vs. halo flattening $q$, having fixed $\beta = 0$, $R_c$ = 5.6 kpc; $v_c(R_0) = 200$\,km/s, $R_0 = 7, 8, 9$\,kpc.}
\end{figure}

Being the relative variations so strong, we do the same analysis as the previous one for every values of $R_0$ and $v_c(R_0)$, evaluating in each case the range for the core radius $R_c$. The results are summarized in table 1.

\begin{table}
\caption{Range for the core radius $R_c$ for models with $\beta = 0$ and different values of $R_0$ and $v_c(R_0)$.}
\begin{center}
\begin{tabular}{|c|c|c|} 
\hline
$R_0$ ( kpc ) & $v_c(R_0)$ ( km/s ) & $R_c$ ( kpc ) \\
\hline
7  &  190  &  12.48 $\div$ 13.86 \\
7  &  200  &  11.37 $\div$ 12.83 \\
7  &  210  &  10.51 $\div$ 11.62 \\
8  &  190  &  15.57 $\div$ 17.55 \\
8  &  200  &  14.03 $\div$ 15.91 \\
8  &  210  &  13.00 $\div$ 14.63 \\
9  &  190  &  18.97 $\div$ 21.91 \\
9  &  200  &  17.24 $\div$ 19.91 \\
9  &  210  &  15.83 $\div$ 18.15 \\
\hline
\end{tabular}
\end{center}
\end{table}

It is worthy to note that the range for $R_c$ never overlaps the observational range and, in one case, exceeds also the 20 kpc permitted by the maximal disk hypothesis. 

Finally, we consider the observational range of $\tau_{obs}$, which is, as we have already pointed out, $1.3 \times 10^{-7} \le \tau_{obs} \le 3.4 \times 10^{-7}$ \cite{B98}, and we do the same analysis for the minimum and the maximum value compatible with errors on $\tau_{obs}$, limiting ourselves to the models with $\beta = 0$, $R_0 = 8$\,kpc, $v_c(R_0) = 200$\,km/s. For $\tau$ equal to $1.3 \times 10^{-7}$ we obtain : 
\begin{equation}
9.05 \ {\rm kpc} \le R_c \le 10.51 \ {\rm kpc} \ ,
\label{eq: sedici}
\end{equation}
which is compatible with the range outlined by observations, while none of our models is compatible with a value of $\tau$ equal to $3.4 \times 10^{-7}$.

\section{Discussion of results for the power\,-\,law models with $\beta = 0$}

We fix now our attention on the results obtained for different power\,-\,law models with $\beta = 0$ and $R_0$ and $v_c(R_0)$ compatible with observations, summarized in table 1 . The ranges for the core radius $R_c$ compatible with the hypothesis of a dark halo entirely composed of MACHO's are given in the third column of the table and we see that they never overlap the observational range of $2 \div 10$\,kpc. We can draw some possible conclusions :

\begin{enumerate}

\item{If we want to keep the hypothesis of axisymmetry, we could take into consideration values for $\beta$ different from zero, i.e. rotation curves which are no more asymptotically flat. In this case, one has to fix a criterium to consider either power\,-\,law models with $\beta > 0$ (declining rotation curve) or $\beta < 0$ (rising rotation curve) and then repeat the analysis present in this paper and then study the results.}

\item{The halo is axisymmetric with flat rotation curves, but the disk is maximal, so that the core radius may be higher than 10 kpc. However, in this case also the contribution to the optical depth of the disk lenses has to be taken into consideration and the extension of the dark halo\footnote{Remember that in this paper we have taken an halo extended till LMC, which is a reasonable hypothesis.} has to be evaluated.}

\item{The dark halo is not entirely composed of MACHO's. The same analysis may be repeated making sections of fig. 1 at different values of $\tau$, choosing the fraction $f$ of the halo in form of MACHO's and imposing that the predicted optical depth $\tau$ for a halo with the specified $f$ is equal to the measured value.}

\end{enumerate}

At the moment it is not easy to choose between the possibilities outlined above. An important point is that, to investigate the shape of the dark halo it is necessary to make observations towards directions other than LMC; in this way more informations on the MACHO's distribution in the Galaxy would be at our disposal.

\section{Separable triaxial models and perfect ellipsoid}

Now, we turn our attention to another class of halo models, which is more general than the axially simmetric we have previously considered, the triaxial halos. Among the triaxial models, the simplest ones belong to the class of the separable triaxial models, a class of self\,-\,gravitating systems whose gravitational potential is of Staeckel form \cite{Staeck90}, \cite{Staeck93}. This kind of potentials are easier to study if we adopt ellipsoidal coordinates, which are defined as follows \cite{MF53} :
\begin{displaymath}
x^2 = \frac{(\lambda + \alpha)(\mu + \alpha)(\nu + \alpha)}
{(\alpha - \beta)(\alpha - \gamma)}
\end{displaymath}
\begin{displaymath}
y^2 = \frac{(\lambda + \beta)(\mu + \beta)(\nu + \beta)}
{(\beta - \alpha)(\beta - \gamma)}
\end{displaymath} 
\begin{displaymath}
z^2 = \frac{(\lambda + \gamma)(\mu + \gamma)(\nu + \gamma)}
{(\gamma - \alpha)(\gamma - \beta)}
\end{displaymath}
where $\alpha$, $\beta$, $\gamma$ are three arbitrary constants. If two of them are equals, ellipsoidal coordinates degenerate in oblate (or prolate) spheroidal coordinates \cite{DeZ85a}, while they reduce to the usual spherical coordinates when $\alpha = \beta = \gamma$.

In the case of a Staeckel potential there are three indipendent integrals of motion \cite{DeZ85a} and these models have been extensively studied in literature. In a classical paper, Kuzmin has studied axysimmetric oblate systems with Staeckel potential and has demonstrated a very useful theorem \cite{K56}, later generalized to generic separable triaxial models by De Zeeuw (1985b). According to this theorem, assigning the mass density profile along the z\,-\,axis is sufficient to determine the whole density profile of a family of models, all having, along the z\,-\,axis, the given mass density profile. Therefore, it is a relatively simple work to construct various models of separable triaxial systems with a Staeckel potential \cite{DPF86}, \cite{DF88}.

We limit ourselves to a class of separable models, called {\em perfect ellipsoids} \cite{DeZ85a}, \cite{DPF86}, \cite{DF88}. We assume the following form for the density profile along the z\,-\,axis :
\begin{equation}
\rho( 0, 0, z) = \psi(z) = \frac{\rho_0 c^n}{(z^2 + c^2)^n}
\label{eq: zaxis}
\end{equation}
being $c$ an arbitrary constant, $\rho_0$ the central density, we consider only the case $n > 0$. Following \cite{DeZ85b}, one may demonstrate that the mass density (in ellipsoidal coordinates) is :
\begin{eqnarray}
\rho(\lambda,\mu,\nu)&=&{g_\lambda}^{2}\psi(\lambda)+{g_\mu}^{2}\psi(\mu)+{g_\nu}^{2}\psi(\nu)+\nonumber\\
& &\nonumber\\
&+& 2g_\lambda g_\mu\frac{\{\Psi(\lambda)-\Psi(\mu)\}}{(\lambda-\mu)}+2g_\mu g_\nu \frac{\{\Psi(\mu)-\Psi(\nu)\}}{(\mu-\nu)}+\nonumber\\
& &\\
&+&2g_\nu g_\lambda\frac{\{\Psi(\nu)-\Psi(\lambda)\}}{(\nu-\lambda)}\nonumber
\label{eq:g_37}
\end{eqnarray}

where $\Psi(\xi)$ $(\xi = \lambda , \mu , \nu)$ is defined as :
\begin{displaymath}
\Psi(\xi) = \int_0^1{\psi(\xi')d\xi'}
\end{displaymath} 
while we have posed :

\begin{equation}
g_\lambda=\frac{(\lambda+\alpha)(\lambda+\beta)}{(\lambda-\mu)(\lambda-\nu)}
\label{eq:g_38}
\end{equation}

\begin{equation}
g_\mu=\frac{(\mu+\alpha)(\mu+\beta)}{(\mu-\nu)(\mu-\lambda)}
\label{eq:g_39}
\end{equation}

\begin{equation}
g_\nu=\frac{(\nu+\alpha)(\nu+\beta)}{(\nu-\lambda)(\nu-\mu)}.
\label{eq:g_40}
\end{equation}

Without loss of generality, we may choose $c^2 = - \gamma$ in (\ref{eq: zaxis}), so eq. (\ref{eq: zaxis}) becomes :
\begin{equation}
\psi(\xi) = \frac{\rho_0 c^n}{\xi^{n/2}}
\label{eq: psi}
\end{equation}
while its primitive is :
\begin{displaymath}
\Psi(\xi) = \frac{2\rho_0 c^2}{2 - n}
\left [
\frac{c^{n-2}}{\xi^{(n-2)/2}}
- 1 \right ]
\ \ \ \ \ \ {\rm for} \ \ n \ne 2 ,
\end{displaymath}
\begin{displaymath}
\Psi(\xi) = \rho_0 c^2 \ln{\frac{\xi}{c^2}} 
\ \ \ \ \ \ {\rm for} \ \ n = 2.
\end{displaymath}

The case of perfect ellipsoid corresponds to choosing $n = 4$ in (\ref{eq: zaxis}); after some calculations, one gets the following expression for the mass density in ellipsoidal coordinates :
\begin{equation}
\rho( \lambda, \mu, \nu) = \rho_0
\left (
\frac{\alpha \beta \gamma}{\lambda \mu \nu}
\right )^2
\label{eq: mass}
\end{equation}

In (\ref{eq: mass}) $\alpha, \beta, \gamma$ are arbitrary constants. Choosing :
\begin{displaymath}
\alpha = - a_1^2 \ \ \ \ \ \
\beta = -a_2^2 \ \ \ \ \ \
\gamma = - c^2 = -a_3^2
\end{displaymath}
and adopting cartesian coordinates, eq. (\ref{eq: mass}) becomes :
\begin{equation}
\rho(\tilde{m}^2) = \frac{\rho_0}{(1 + \tilde{m}^2)^2} =
\frac{\rho_{\odot} (a_2^2 + R_0^2)^2}{a_2^4 ( 1 + \tilde{m}^2)^2}.
\label{eq: tsette}
\end{equation}

Here, $\rho_{\odot}$ and $R_0$ are respectively the local mass density and the distance of the Sun to the galactic centre and :
\begin{equation}
\tilde{m}^2 = \frac{x^2}{a_1^2} + 
\frac{y^2}{a_2^2} + \frac{z^2}{a_3^2}.
\label{eq: totto}
\end{equation}

In (\ref{eq: totto}) $a_1, a_2 \ {\rm and} \ a_3$ are the semiaxes of the concentric equipotential ellipsoidal surfaces. They must satisfy the relation $a_1 > a_2 > a_3$, but their exact values are not known. It is interesting to note that this is the only separable triaxial model whose equipotential surfaces are exactly similar to concentric ellipsoids \cite{DL85}. The mass inside the ellipsoidal radius is :
\begin{equation}
M_{ell} = 2\pi \rho_0 a_1 a_2 a_3 
\left ( 
\arctan{\tilde{m}} - \frac{\tilde{m}}{1 + \tilde{m}}
\right ) .
\label{eq: tnove}
\end{equation}
The total mass is finite and is given by :
\begin{equation}
M_{tot} = \pi^2 \rho_0 a_1 a_2 a_3 = 
\pi^2 \rho_{\odot} a_1 a_2 a_3 ( 1 + R_0^2/a_2^2 )^2 .
\label{eq: tdieci}
\end{equation}

The DF of this model is not known. Some authors \cite{DeL91} have recently tried to reconstruct the DF of separable triaxial models, looking for a formula similar to the Eddington one for spherical models \cite{BT87}. The result is very difficult to obtain and it is not unique, this is the reason why we cannot calculate the event rate, therefore we limit ourselves to evaluate only the optical depth. We do it in the following section.

\section{Optical depth towards LMC and estimate of axial ratios}

Evaluating optical depth through the formula (\ref{eq: quattro}) requires the mass density of the model, which is given, for the perfect ellipsoid, by equation (\ref{eq: tsette}). However, we have first to change the coordinates, from galactocentric to heliocentric and then use the formula (\ref{eq: quattro}). The result is :
\begin{equation}
\tau = \frac{4 \pi G D_s^2}{c^2} a_1^4 a_3^4 (a_2^2 + R_0^2)^2 I
\label{eq: tundici}
\end{equation}
where $I$ is the integral :
\begin{equation}
I = \int_0^1{\frac{s (1 - s)}{(As^2 + Bs + C)^2} ds}
\label{eq: tdodici}
\end{equation}
This may be calculated analitically. We get :
\begin{equation}
I = \frac{2}{\Delta} + \frac{2B + 4C}{\Delta ^{3/2}}
\left [
\arctan{\frac{b}{\sqrt{\Delta}}} - \arctan{\frac{B + 2A}{\sqrt{\Delta}}}
\right ]
\label{eq: ttredici}
\end{equation}
with :
\begin{displaymath}
A = D_s^2(a_2^2 a_3^2 \sin^2{b} \cos^2{l} + a_1^2 a_3^2 \sin^2{b} \sin^2{l} +
a_1^2 a_2^2 \cos^2{b}) ;
\end{displaymath}
\begin{displaymath}
B = 2D_s R_0 a_1^2 a_3^2 \sin{b} \sin{l} ;
\end{displaymath}
\begin{displaymath}
C = a_1^2 a_3^2 R_0^2 + a_1^2 a_2^2 a_3^2 ;
\end{displaymath}
\begin{displaymath}
\Delta = 4AC - B^2 > 0.
\end{displaymath}

Now, to calculate the optical depth towards LMC, we have only to insert in (\ref{eq: tundici}) and (\ref{eq: ttredici}) the values $(b, l) = (-32^{\circ}.9, 280^{\circ}.5)$ and $D_s = 50$\,kpc; we get the expression of $\tau$ in terms of the semiaxes of the perfect ellipsoid. We apply the method of the inverse problem to estimate the axial ratios of a perfect ellipsoidal dark halo for which the predicted optical depth is equal to the observed one. However, we have to fix the value of the halo extension, i.e. we have to fix $D = a_1$. This latter quantity is not well determined and different authors give different estimates of $D$, in the range $(50 \div 200) \ {\rm kpc}$ (for a brief review, see \cite{Z98}). A possible choice is $D = 100 \ {\rm kpc}$. In fig. 4a we plot the optical depth towards LMC as function of axial ratios $a_2/a_1$, $a_3/a_1$, while fig. 4b is a section of 4a at the level $\tau = 2.1 \times 10^{-7}$.

\begin{figure}
\centering
\mbox{ \epsfxsize=13cm \epsfysize=10cm \epsffile{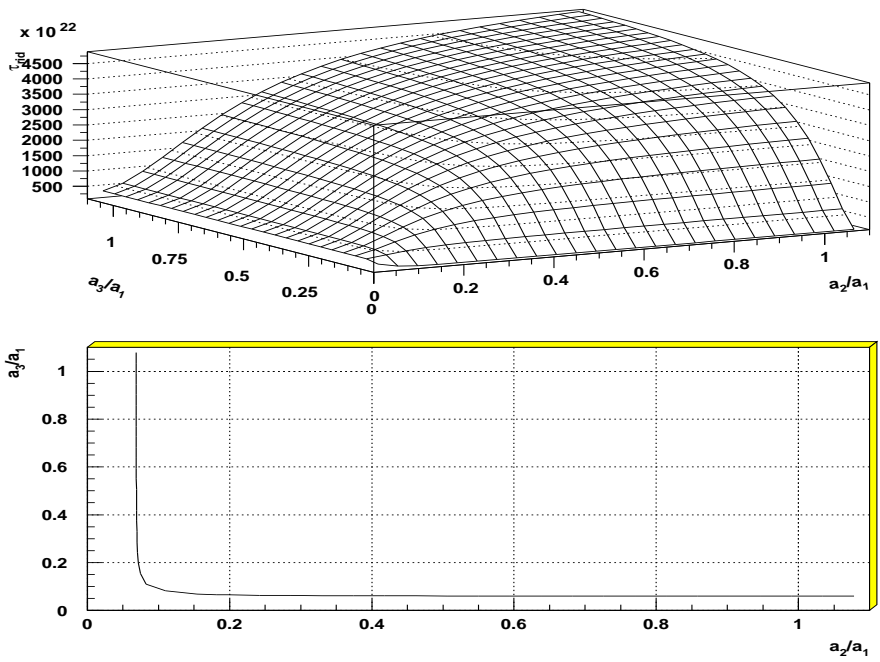}}
\caption{a) Optical depth $\tau$ \ towards LMC vs. \ $a_2/a_1$ \ and \ $a_3/a_1$ for the perfect ellipsoid, having fixed the halo extension as $a_1 = D = 100$\,kpc. On the z\,-\,axis there is the quantity $\tau_{rid} = \tau/G\rho_{\odot}$. b) Section of a) for $\tau = 2.1 \times 10^{-7}$.}
\end{figure} 

\begin{table}
\caption{Values of $a_2/a1$ and $a_3/a_1$ to have a perfect ellipsoidal dark halo which predicts an optical depth towards LMC equal to what is measured.}
\begin{center}
\begin{tabular}{|c|c|} 
\hline
$a_2/a_1$ & $a_3/a_1$ \\
\hline
0.2  &  0.064 \\
0.3  &  0.063 \\
0.4  &  0.062 \\
0.5  &  0.061 \\
0.6  &  0.061 \\
0.7  &  0.061 \\
0.8  &  0.061 \\
0.9  &  0.061 \\
1.0  &  0.061 \\
\hline
\end{tabular}
\end{center}
\end{table} 

From fig. 4b one can see that for each value of $a_2/a_1$ there is only one value of $a_3/a_1$ for which the predicted optical depth equals the observed one. Table 2 shows how $a_3/a_1$ changes varying $a_2/a_1$. From the condition on $a_1,~ a_2,~ a_3$ it has to be $a_3/a_1< a_2/a_1$; from fig. 4b this happens when $a_2/a_1> 0.1$, and the value of $a_3/a_1$ assumes an approximately constant value. A good value is then $a_3/a_1 \simeq constant = 0.06$. 

Now we should determine the other axial ratio, $a_2/a_1$. We could use the formula (\ref{eq: tdieci}) which gives the total mass of the halo, rewritten as :
\begin{equation}
M_{tot} = \pi^2 \rho_{\odot} a_1^3 (a_3/a_1)
\frac{(R_0^2/a_1^2 + a_2^2/a_1^2)^2}{(a_2/a_1)^4}
\label{eq: masstot}
\end{equation}
and determine the ratio $a_2/a_1$ fixing the total mass of the halo. 

\begin{figure}
\centering
\mbox{ \epsfxsize=13cm \epsfysize=10cm \epsffile{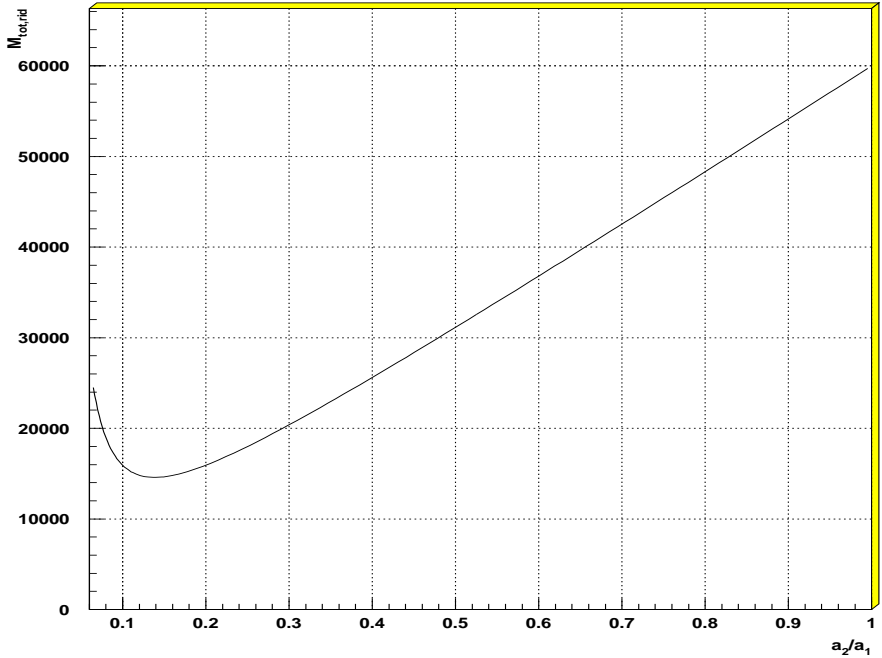}}
\caption{The total mass of the perfect elliposid in function of $a_2/a_1$, having fixed $a_1 = 100$\,kpc and $a_3/a_1 = 0.06$; only for $a_2/a_1> 0.1$ have to be taken into consideration. On the y\,-\,axis there is $M_{tot,rid} = M_{tot}/\rho_{\odot}$.}
\end{figure}

This latter is known with great uncertainties and different estimates are present, which adopt different values for the halo extension and a great variety of method and models (see e.g. \cite{L95}, \cite{Z98}). We could use one of the estimates to obtain the corresponding value of $a_2/a_1$, but we prefer just to point out the possibility of estimating the total mass of the dark halo, once the ratio $a_2/a_1$ is known through other ways. Fig. 5 is a plot of $M_{tot} = M_{tot}(a_2/a_1)$, and only for $a_2/a_1> 0.1$ have to be taken into consideration.

\section{Discussion of results for the perfect ellipsoid}

Assuming a 100\% MACHO's halo, we have analysed what are the features of a perfect ellipsoidal dark halo to have the theoretical optical depth towards LMC compatible with the observational values. We have obtained an estimate for the ratio $a_3/a_1$, and expressing the total mass of the dark halo as a function of the axial ratios, we have the possibility, either to estimate he mass of the dark halo, fixing a value for $a_2/a_1$, or the axial ratio $a_2/a_1$ once the total mass is known with a good degree of accuracy. The halo extension is taken as $D$ = 100 kpc.

The value we obtained for the axial ratio $a_3/a_1 \simeq 0.06$ shows an extremely flattened dark halo in the direction perpendicular to the disk plane. This result is consistent with the predictions of numerical simulations with dissipational dark matter \cite{D94}, \cite{R95}. Ostriker \& Peebles (1973) have studied the stability of flattened systems, determining a critical value for the ratio between kinetic and potential energy, above which the system is unstable against bar formation. They studied different models, but all of them are axisymmetric. This is not the case we are considering, so their results should be generalized in order to  be applied. Few evidences are in favour of the presence of a bar in the bulge of our Galaxy \cite{blitz}, \cite{dwek}, some come from microlensing observations \cite{zhao1}, \cite{zhao2}, but it is relative to the bulge component, and it is not easy to evaluate whether there is a significant contribution due to the halo, which is usually considered negligible in that region of the Galaxy.

\section{Conclusions}

In this paper we have seen how it is possible to infer the properties of the dark halo of our Galaxy from microlensing observations: what we have called the {\it inverse problem in microlensing}. This method of investigation of the dark halo has been applied to two different halo models, the power\,-\,law models with flat rotation curves and the perfect ellipsoid model.

First, we have shown how the observed optical depth may restrict the range for the core radius for power\,-\,law models with flat rotation curves, having supposed a 100\% MACHO's halo. The values we found are always higher than the observational ones, which are in the range $(2 \div 10) \ {\rm kpc}$, as delineated by the work of Bachall, Schmidt \& Soneira and Caldwell \& Ostriker. These results may be interpreted in different ways, as we pointed out in Section 5, and it is worthy to note that, when added to other ones coming from astronomical observations, our results give some indications on the validity of the description of the dark halo through power\,-\,law models. 

Next, we have considered a triaxial shape for the dark halo, adopting the perfect ellipsoid model, a simple separable triaxial system. Describing the halo as a triaxial structure is a difficult task since there are not enough observational constraints in order to determine all the properties of the model. However, fixing the halo extension, we have been able to determine one of the axial ratios ($a_3/a_1$) and find a relation between the total mass of the halo and the other axial ratio ($a_2/a_1$). 

We have applied the method of the inverse problem to constraint the dark halo on the basis of microlensing observations only in one direction, the LMC, and in fact we could not determine uniquely the properties of the models we considered. In order to investigate the shape of the halo and put more precise constraints on the relative parameters, it will be necessary to have microlensing data relative to observations towards other directions, like globular clusters, for example; they may be used as sources (see e.g. \cite{GH97}) or as sites of lenses when observing towards SMC (see e.g. \cite{JSW98}). The same method can be used, of course, to investigate other halo shapes.

It is a great pleasure to thank G. Busarello and Ph. Jetzer for the discussion we had on the manuscript. 

\bibliography{invprob}
\end{document}